\documentclass[prl]{revtex4}
\usepackage{graphicx}

\input epsf

\newcommand{\ket}[1]{|#1\rangle}

\newcommand{\brak}[1]{\langle#1|}

\newcommand{\be}[1]{\begin{equation}}

\newcommand{\ee}[1]{\end{equation}}

\begin{document}

\title{Phase-dependent interaction in a 4-level atomic configuration}

\author{Giovanna Morigi$^{1*}$, Sonja Franke-Arnold$^2$, and Gian-Luca
Oppo$^2$}

\address{$^1$Max-Planck-Institut f\"ur Quantenoptik, Hans-Kopfermannstr. 1,
D-85748 Garching, Germany \\
$^2$ Dept.~of Physics and Applied
Physics, University of Strathclyde, 107 Rottenrow, Glasgow, G4 0NG
United Kingdom}

\date{\today}

\begin{abstract}

\noindent We study a four-level atomic scheme interacting with four lasers in a
closed-loop configuration with a $\diamondsuit$ (diamond) geometry. We
investigate the influence of the laser phases on the steady state. We show
that, depending on the phases and the decay characteristic, the system can
exhibit a variety of behaviors, including population inversion and complete
depletion of an atomic state. We explain the phenomena in terms of multi-photon
interference. We compare our results with the phase-dependent phenomena in the
double-$\Lambda$ scheme, as studied in [Korsunsky and Kosachiov, Phys. Rev A
{\bf 60}, 4996 (1999)]. This investigation may be useful for developing
non-linear optical devices, and for the spectroscopy and laser-cooling of
alkali-earth atoms.
\end{abstract}

\maketitle

\section{Introduction}

Absorption and emission of monochromatic light in two-level atomic transitions
are well-understood processes in quantum optics \cite{Eberly}. Their
properties, however, can change drastically if transitions to a third atomic
level have to be included.  This is the case, for instance, in the $\Lambda$
configuration where two stable states are coupled to a common excited state by
laser fields, thus providing two excitation paths which can interfere. This
interference lies at the heart of Coherent Population Trapping (CPT)
\cite{CPT}. Here, destructive interference between the transition amplitudes
gives rise to a superposition of atomic states (dark state) that is
decoupled from coherent radiation but populated by spontaneous emission.
Consequently the atom becomes ``trapped'' in this coherent superposition.

For configurations like the $\Lambda$ scheme, the relative phase of the laser
fields does not
affect the steady-state dynamics, in the sense that there always exists a
reference frame in which the Rabi frequencies are real. This is no longer
fulfilled in closed-loop configurations \cite{Buckle86,Kosachiov92}, i.e. when
a set of atomic states is (quasi- ) resonantly coupled by laser fields,
such that each state of the set is connected to any other via two different
paths of coherent photon-scattering: In this case, the relative phase $\Phi$
between the transitions determines the interference and hence critically
influences the dynamics and steady state of the system
\cite{Buckle86,Kosachiov92,Korsunsky99}. Previous studies of closed-loop
configurations often featured double-$\Lambda$ systems, where two stable or
metastable states are -each- coupled to two common excited states
\cite{Korsunsky99,PhaseoniumRev,Windholz96,Harris00}. These works have shown a
rich variety of non-linear optical phenomena.

In this paper, we investigate the phase-dependent dynamics of a closed-loop
configuration, consisting of four transitions driven by lasers. One
ground state is coupled in a V-type structure to two intermediate states, which
are themselves coupled to a common excited state in a $\Lambda$-type structure.
We label this system
the $\diamondsuit$ (diamond) scheme. The steady-state of the $\diamondsuit$
scheme, like that of the double-$\Lambda$ configuration, is a periodic function
of the relative phase $\Phi$ between the excitation paths, which contribute to
the scattering between any initial and final state of the scheme. In
particular, the steady state is determined by the concurrence of the
phase-dependent Hamiltonian dynamics and the relaxation processes. We will show
that, depending on $\Phi$ and on the lifetimes of the intermediate states, the
$\diamondsuit$ system can show a variety of behaviors including population
inversion, CPT, and phase-dependent refractive indices. It is worth noting that
the double-$\Lambda$ and the $\diamondsuit$ schemes are governed by the same
Hamiltonian, but are characterized by different relaxation processes. This
results in critical differences in the dynamics, which we will point out in our
discussion.

Excitation configurations like the $\diamondsuit$ scheme have been investigated
in the literature as a model for observing pressure-induced resonances
\cite{Grynberg90}, and can be found for instance in experiments with gases of
alkali-earth atoms which aim at
optical frequency standards \cite{PTB01} or at reaching
the quantum degeneracy regime by
all-optical means \cite{Katori99}. Our investigation may contribute to the
spectroscopy of these systems and to the development of new and efficient
methods of laser-cooling.

This paper is organized as follows. In section II we introduce the model and
the basic equations. We discuss the properties of the system,
and identify some relevant parameter regimes.
In section III we calculate the steady-state solutions as a function of the
relative phase between excitation paths for certain parameters, and discuss the
results. In Section IV we draw the conclusions and discuss some outlooks.

\section{The Model}

In this section we introduce the model, using a density-matrix formalism. We
move then to a reference frame where the phase dependence is explicit, and
formulate the optical Bloch equations. Finally, we discuss the symmetries of
the system as a function of the phase and of the relaxation processes.

\subsection{The Master equation for a single atom}

We consider a gas of atoms of mass $m$. The atoms are free, and interact with a
multi-chromatic light field. For a sufficiently-dilute gas, each atom interacts
individually with the light, which couples to a set of atomic levels as
depicted in Fig.~1. This set contains a ground state $|1\rangle$, two
intermediate states $|2\rangle$ and $|3\rangle$ and an excited state
$|4\rangle$. The transitions $|1\rangle \to |2\rangle, |3\rangle$ and
$|2\rangle,|3\rangle\to |4\rangle$ are optical dipoles with decay rates
$\gamma_2$, $\gamma_3$, $\gamma_{42}$, and $\gamma_{43}$, respectively. Each
dipole transition is driven resonantly by a laser, which is here considered to
be a classical running wave, propagating along the $\hat{z}$ axis. The laser
coupling to the transition $|i\rangle \to |j\rangle$ is characterized by
frequency $\omega_{ij}$ and wave vector $k_{ij}$, while the strength of the
coupling is given by the Rabi frequency $g_{ij}{\rm e}^{{\rm i}\chi_{ij}}$,
where $g_{ij}$ is real and $\chi_{ij}$ is a constant phase, determined by the
phase of the atomic dipole and by the phase of the laser at time $t=0$ and
position $z=0$. The state of one atom at time $t$ is described by the density
matrix $\sigma$, which obeys the Master equation:
\begin{equation}
\label{Master:0} \frac{\partial}{\partial t}\sigma=\frac{1}{{\rm
i}\hbar}\left[H(t),\sigma\right] +{\cal L}\sigma.
\end{equation}
Here, the Hamiltonian $H(t)$ contains the coherent dynamics of the atom and the
Liouvillian ${\cal L}$ describes the relaxation processes. The Hamiltonian
depends explicitly on time and reads:
\begin{eqnarray}
\label{Hamiltonian} H(t) &=& \frac{p_z^2}{2m}+\sum_{j=2}^4 \hbar \omega_j
\ket{j}\brak{j} \nonumber \\ & & +\frac{\hbar}{2} \sum_{j=2,3}\bigl( g_{1j}{\rm
e}^{-i(\omega_{1j} t-k_{1j}z+\chi_{1j})}\ket{j}\brak{1} \nonumber  \\ & & +
g_{j4}{\rm e}^{-i(\omega_{j4} t-k_{j4}z+\chi_{j4})}\ket{4}\brak{j} +{\rm
H.c.}\bigr).
\end{eqnarray}
The first term corresponds to the kinetic energy of the atomic center of mass,
where $p_z$ is its momentum along the $\hat{z}$ axis (for simplicity, we
consider only the component of the atomic motion along the $\hat{z}$ axis). The
second term corresponds to the internal Hamiltonian, where $\hbar\omega_j$
denote the energies of the atomic states $\ket{j}$ relative to the energy of
the state $|1\rangle$. The remaining term describes the atom-laser interaction.

\noindent The relaxation processes are assumed to be solely radiative, and
the Liouvillian ${\cal L}$ in (\ref{Master:0}) has the form:
\begin{eqnarray}
\label{Eq:2} {\cal L}\sigma &=&\sum_{j=2,3}\gamma_{4j} \int_{-1}^{1} {\rm d}u
{\cal N}_{4j}(u) {\rm e}^{-{\rm i}k_{j4}uz}|j\rangle\langle 4|\sigma
|4\rangle\langle j| {\rm e}^{{\rm i}k_{j4}uz}
-\frac{\gamma_4}{2}\left(|4\rangle\langle 4|\sigma +\sigma |4\rangle\langle
4|\right)\\ & & +\sum_{j=2,3} \gamma_j\int_{-1}^1 {\rm d}u {\cal N}_{j1}(u)
\left[{\rm e}^{-{\rm i}k_{1j}uz}|1\rangle\langle j|\sigma |j\rangle\langle 1|
{\rm e}^{{\rm i}k_{1j}uz} -\frac{1}{2}\left(|j\rangle\langle j|\sigma +\sigma
|j\rangle\langle j|\right)\right]\nonumber,
\end{eqnarray}
where $\gamma_4=\gamma_{42}+\gamma_{43}$ is the total decay rate from level
$\ket{4}$, and ${\cal N}_{ij}(u)$ is the dipole pattern for the spontaneous
emission of a photon on the transition $|i\rangle\to |j\rangle \;(j=2,3,4)$,
i.e. the probability of emission at the angle $\theta$ from the $\hat{z}$ axis,
such that $u=\cos\theta$. This term describes the diffusion in the dynamics of
the atomic motion due to the randomness of the incoherent events. In the limit
in which the center-of-mass motion can be treated classically, and for
sufficiently short times this effect can be neglected so that
the explicit form of the dipole pattern is irrelevant. On this time scale
and for a dilute gas, we can neglect relaxations due to collisions
between the atoms. This is the regime that we are considering in
the main body of the paper.

\subsection{Change of reference frame}

The Hamiltonian (\ref{Master:0}) is explicitly time-dependent. For the
configuration we consider here there exists in general no reference frame in
which this explicit time-dependence can be eliminated. This is a characteristic
of "closed-loop" configurations \cite{Kosachiov92}, and it is a manifestation
of the intrinsic phase sensitivity of the dynamics. However, in an adequate
reference frame, the Rabi frequencies can be chosen such that only one is
complex, with its phase $\Phi$ being a function of all laser phases.
Without loss of generality, we move to a reference frame where $\Phi$ is
associated with the laser coupling to the transition $|3\rangle\to |4\rangle$,
so that the coherent dynamics of the system is now described by the
Hamiltonian \cite{Buckle86}
\begin{eqnarray}
\label{H:new} \tilde{H}(\Phi) &=&\frac{p_z^2}{2m}+\sum_{j=2}^4 \hbar \delta_{j}
\ket{j}\brak{j}\nonumber\\ & & + \frac{\hbar}{2}
\bigl(g_{12}\ket{2}\brak{1}+g_{13}\ket{3}\brak{1}+g_{24}
\ket{4}\brak{2}+g_{34}{\rm e}^{{\rm i}\Phi}\ket{4}\brak{3}+{\rm H.c.}\bigr),
\end{eqnarray}
where {\bf $\Phi=\Phi(t,z)$}. The
new Hamiltonian is connected with $H(t)$ by the relation:
$\tilde{H}(\Phi)=U^{-1}H(t)U-{\rm i}\hbar U^{-1}\partial U/\partial t$, where
$U(t)$ is a unitary transformation, which reads
\begin{eqnarray}
U(t)
&=&\exp\left\{-{\rm i} \left[(\omega_{12}+\omega_{24})t
  -(k_{12}+k_{24})z+(\chi_{12}+\chi_{24})\right]
|4\rangle\langle 4| \right\}\\
&\times &\exp\left\{-{\rm i}
\left[(\omega_{12}t-k_{12}z+\chi_{12}) |2\rangle\langle 2|
+(\omega_{13}t-k_{13}z+\chi_{13}) |3\rangle\langle 3|\right]\right\}.
\nonumber
\end{eqnarray}
The detunings $\delta_j$ in (\ref{H:new}) are given by
\begin{eqnarray*}
\delta_2&=&(\omega_2-\omega_{12})+\frac{k_{12}p_z}{m} +\frac{\hbar
k_{12}^2}{2m},\\ \delta_3&=&(\omega_3-\omega_{13})
+\frac{k_{13}p_z}{m}+\frac{\hbar k_{13}^2}{2m},\\
\delta_4&=&(\omega_4-\omega_{12}-\omega_{24})
+\frac{(k_{12}+k_{24})p_z}{m}+\frac{\hbar(k_{12}+k_{24})^2}{2m},
\end{eqnarray*}
and the phase $\Phi$ is defined as
\begin{equation}\label{eq:phase}
\Phi=\Delta \omega t-\Delta k z + \Delta \chi,
\end{equation}
where
\begin{eqnarray}
\Delta\omega
& \doteq& \omega_{12}+\omega_{24}-\omega_{13}-\omega_{34},\\
\Delta k
& \doteq & k_{12}+ k_{24}- k_{13}-k_{34},\\
\Delta\chi
& \doteq & \chi_{12}+ \chi_{24}- \chi_{13}-\chi_{34}.
\end{eqnarray}
The phase $\Phi$ is the relative phase between the two excitation paths
characterizing any transition between two atomic states. This phase is in
general time- and position-dependent, and it results from the multi-photon
detuning $\Delta\omega$, the wave-vector mismatch $\Delta k$, and from the
initial laser and atomic-dipole phases, $\Delta \chi$.\\ In the new reference
frame the density matrix is $\rho=U^{-1}(t)\sigma U(t)$,
and its evolution is described by the Master Equation
\begin{equation}
\label{Master:new}
\frac{\partial}{\partial t}\rho=\frac{1}{{\rm
i}\hbar}\left[\tilde{H}(\Phi),\rho\right] +{\cal L}\rho \, .
\end{equation}
In the next section we derive the equations of motion for the matrix
elements of $\rho$.

\subsection{Optical Bloch Equations}

In the following we assume a thermal distribution of atoms at temperature $T$,
such that the thermal energy $k_BT$ is much larger than the recoil energies
$\hbar^2 k^2_{ij}/2m$. In this limit, we treat the atomic motion classically,
and neglect the effects of the photon recoil on the center-of-mass dynamics
\cite{Footnote}. Then, the one-atom density matrix is given by $\rho=\int {\rm
d}z{\rm d}p_z n(z,p_z)\rho(p_z,z)$, where $n(z,p_z)$ is the atomic density as a
function of the position and the momentum, and $\rho(z,p_z)$ obeys the Master
equation (\ref{Master:new}) at the parameters $z,p_z$. Denoting with
$\rho_{ij}=\langle i|\rho(z,p_z)|j\rangle$ its elements, with $i,j=1,2,3,4$
(we drop the explicit dependence on the parameters $(z,p_z)$~), the Optical
Bloch Equations (OBE) have the form:
\begin{eqnarray}
\label{Bloch1}
\dot{\rho}_{11}
&=& {\rm i} \frac{g_{12}}{2}(\rho_{12}-\rho_{21})
  + {\rm i} \frac{g_{13}}{2}(
\rho_{13}-\rho_{31})\nonumber\\ & & +\gamma_2\rho_{22}+\gamma_3\rho_{33},\\
\dot{\rho}_{22}
&=& -{\rm i} \frac{g_{12}}{2}(\rho_{12}-\rho_{21})
   + {\rm i} \frac{g_{24}}{2}(\rho_{24}-\rho_{42}) \nonumber\\
& & + \gamma_{42}\rho_{44}-\gamma_2\rho_{22},\\
\dot{\rho}_{33} &=&  -{\rm i}\frac{g_{13}}{2} (\rho_{13}-\rho_{31})
    + {\rm i}\frac{g_{34}}{2}({\rm e}^{{\rm i}\Phi}\rho_{34}-{\rm e}^{-{\rm i}\Phi}\rho_{43})\nonumber\\
& & +\gamma_{43}\rho_{44}-\gamma_3\rho_{33},\\
\rho_{44}
&=& 1-\rho_{11}-\rho_{22}-\rho_{33},\\
\dot{\rho}_{12} &=& \left({\rm
i}\delta_2-\frac{\gamma_2}{2}\right)\rho_{12} -{\rm
i}\frac{g_{12}}{2}(\rho_{22}-\rho_{11})\nonumber\\ & & -{\rm
i}\frac{g_{13}}{2}\rho_{32} +{\rm i}\frac{g_{24}}{2}\rho_{14},\\
\dot{\rho}_{13} &=&\left({\rm
i}\delta_3-\frac{\gamma_3}{2}\right)\rho_{13}
         -{\rm i}\frac{g_{13}}{2}
(\rho_{33}-\rho_{11})\nonumber\\ & & -{\rm i}\frac{g_{12}}{2}\rho_{23}+{\rm
i}\frac{g_{34}}{2}{\rm e}^{{\rm i}\Phi}\rho_{14},\\
\dot{\rho}_{24}
&=&\left(-{\rm i}(\delta_{2}-\delta_4)-
\frac{\gamma_2+\gamma_4}{2}\right)\rho_{24}
         -{\rm i}\frac{g_{24}}{2}(\rho_{44}-\rho_{22})\nonumber\\
& & -{\rm i}\frac{g_{12}}{2}\rho_{14} +{\rm i}\frac{g_{34}}{2}{\rm e}^{-{\rm
i}\Phi}\rho_{23},\\
\dot{\rho}_{34} &=&\left(-{\rm
i}(\delta_{3}-\delta_4)-\frac{\gamma_3+\gamma_4}{2}\right) \rho_{34} -{\rm
i}\frac{g_{34}}{2}{\rm e}^{-{\rm i}\Phi}(\rho_{44}-\rho_{33})\nonumber\\ & &
-{\rm i}\frac{g_{13}}{2}\rho_{14}
    +{\rm i}\frac{g_{24}}{2}\rho_{32},\\
\dot{\rho}_{14} &=&\left[{\rm i}\delta_{4}
-\frac{\gamma_4}{2}\right]\rho_{14} -{\rm
i}\frac{g_{12}}{2}\rho_{24}-{\rm i}\frac{g_{13}}{2}
\rho_{34}\nonumber\\ & & +{\rm i}\frac{g_{24}}{2}\rho_{12}
+{\rm i}\frac{g_{34}}{2}{\rm e}^{-{\rm i}\Phi}\rho_{13},\\
\label{BlochN} \dot{\rho}_{23} &=&\left[-{\rm
i}\left(\delta_{2}-\delta_{3}\right)
-\frac{\gamma_2+\gamma_3}{2}\right]\rho_{23} -{\rm
i}\frac{g_{12}}{2}\rho_{13}\nonumber\\ & & -{\rm i}\frac{g_{24}}{2}\rho_{43}
+{\rm i}\frac{g_{13}}{2}\rho_{21} +{\rm i}\frac{g_{34}}{2}{\rm e}^{{\rm
i}\Phi}\rho_{24},
\end{eqnarray}
and $\rho_{ji}=(\rho_{ij})^*$. In the following we will refer to the
diagonal elements as ``populations", giving the occupation of the atomic
states, and to the off-diagonal elements as ``one-photon coherences" or
``two-photon coherences", depending on whether the involved states are coupled
at lowest order by the scattering of one or two photons, respectively.

\subsubsection{Discussion}

Equations (\ref{Bloch1}-\ref{BlochN}) exhibit a parametric time dependence,
which enters through the phase $\Phi$ as in (\ref{eq:phase}). Such behavior
imposes limitations on the existence of a steady-state solution. Neglecting the
coupling of the internal degrees of freedom with the external ones, two cases
can be identified where the internal steady state exists: (i) when the scheme
is driven well below saturation, and/or (ii) for $\Delta\omega=0$. In (i), the
processes leading to the absorption of two photons are negligible, and the
relevant dynamics takes place between the ground state $\ket{1}$ and the
intermediate states $\ket{2},\ket{3}$, coupled in a (open-loop) V-type
configuration \cite{Footnote2}. In (ii), the steady-state solution is defined
for any value of the other parameters. For $\Delta\omega=0$ and $\Delta k=0$,
$\Phi$ is determined solely by the initial laser- and dipole-phases. While the
dipole phases are fixed by the quantum numbers of the atomic transitions, the
laser phases can be modified. In this regime, which we study in the remainder
of the paper, the dynamics of the $\diamondsuit$ configuration shares some
analogies with an interferometer, the arms of which are formed by the
multiphoton excitation paths and the phase difference between them is given by
the relative phase $\Phi$. Such analogy was first drawn in \cite{Buckle86} for
the coherent dynamics of a closed-loop scheme. In this spirit we interpret some
of the results, presented below and in the next section, which have been
obtained in the presence of relaxation processes.\\ An interesting
manifestation of the phase-dependent dynamics is the probability of two-photon
absorption on the transition $\ket{1}\to\ket{4}$. In the limit of weak
excitations it has the form
\begin{equation}
P_{1\to 4}\propto\Bigl|\frac{g_{12}g_{24}}{\delta_{2} -{\rm
i}\gamma_2/2} +\frac{g_{13}g_{34}\exp({\rm i}\Phi)}{\delta_{3}
-{\rm i}\gamma_3/2}\Bigr|^2 . \label{Phase:Pi}
\end{equation}
Here, the first and second terms on the RHS describe the transitions via the
intermediate states $\ket{2}$ and $\ket{3}$, respectively. From
(\ref{Phase:Pi}) it is evident that both the phase difference $\Phi$ as well as
the ratio of the laser detunings $\delta_{2},\delta_{3}$ and the decay rates
$\gamma_2,\gamma_3$ determine the interference between the two paths. In the
special case of equal Rabi frequencies $g_{ij}=g$, and for
$\delta_{2}=\delta_{3}$, $\gamma_2=\gamma_3$, the role of $\Phi$ is singled
out{\bf :}
\begin{equation}
\label{P1to4} P_{1\to 4}\propto \cos^2\frac{\Phi}{2} \; .
\end{equation}
Thus, the transition probability from the ground to the excited level is
modulated by $\Phi$. In particular, it is maximal for the values $\Phi=2n\pi$
(where $n$ is an integer), while it vanishes for $\Phi=(2n+1)\pi$. At the
latter value, no transition to $\ket{4}$ occurs. In the next section we will
show that $P_{1\to 4}$ always vanishes for $\Phi=(2n+1)\pi$ at steady state,
even when the system is driven at saturation. We remark that the appearance of
this behavior requires a "symmetric" excitation configuration, meaning that
each two-photon excitation path from $|1\rangle$ to $|4\rangle$ has,
separately, the same probability.\\

A further understanding of the problem can be gained by moving to a suitable
basis, following the analysis of \cite{Buckle86}. This basis is chosen
appropriate to the structure of the Hamiltonian (\ref{H:new}) and the
relaxation processes in (\ref{Eq:2}). For $g_{ij}=g$ and $\delta_j=0$ the
dynamics offers simple interpretations for $\Phi=n\pi$.

We first focus on the values $\Phi=(2n+1)\pi$. Here, it is convenient to use
the orthogonal basis set
$\{\ket{1},\ket{4},\ket{\Psi_{23}(0)},\ket{\Psi_{23}(\pi)}\}$, where
\begin{equation}
\ket{\Psi_{23}(\theta)}=\frac{1}{\sqrt{2}}
\left[\ket{2}+{\rm e}^{-{\rm i}\theta}\ket{3}\right]
\end{equation}
with $\theta=0,\pi$. In this basis, the Hamiltonian (\ref{H:new}) can be
rewritten as
\begin{equation}
\label{H:Phi23} \tilde{H}((2n+1)\pi)=\frac{\hbar g}{\sqrt{2}}
\Bigl[\ket{\Psi_{23}(0)}\brak{1}+
\ket{4}\brak{\Psi_{23}(\pi)}+{\rm H.c.}\Bigr],
\end{equation}
where we have omitted the atomic motion. Thus, (\ref{H:Phi23}) describes
two-level dynamics within the orthogonal subspaces
$\{\ket{1},\ket{\Psi_{23}(0)}\}$ and $\{\ket{4},\ket{\Psi_{23}(\pi)}\}$. These
subspaces are coupled by spontaneous decay, and the coupling between states due
to coherent and incoherent processes is represented in Fig.~2 a). From the
structure of the decay, it is evident that the atom is eventually pumped into
$\{\ket{1},\ket{\Psi_{23}(0)}\}$. Hence, the steady state of the system for
this value of the phase corresponds to that of the driven two-level transition
$\ket{1} \to \ket{\Psi_{23}(0)}$.\\ \indent For $\Phi=2n\pi$ we describe the
system in the orthogonal basis set
$\{\ket{\Psi_{14}(0)},\ket{\Psi_{14}(\pi)},\ket{\Psi_{23}(0)},
\ket{\Psi_{23}(\pi)}\}$, where
\begin{equation}
\ket{\Psi_{14}(\theta)}=\frac{1}{\sqrt{2}} \left[\ket{1}+{\rm
e}^{{\rm i}\theta}\ket{4}\right]\\
\end{equation}
with $\theta=0,\pi$. In this basis, the Hamiltonian $\tilde{H}$ can be written
as
\begin{equation}
\label{H:Phi14}
\tilde{H}(2n\pi)=\hbar g
\Bigl[\ket{\Psi_{23}(0)}\brak{\Psi_{14}(0)}+{\rm H.c.}\Bigr],
\end{equation}
and it describes a coherent two-level dynamics between the states
$\ket{\Psi_{23}(0)}$ and $\ket{\Psi_{14}(0)}$. The states
$\ket{\Psi_{14}(\pi)}$ and $\ket{\Psi_{23}(\pi)}$ are decoupled from the
coherent drive because of destructive interference between the
corresponding excitation paths, and from this point of view they are dark
states. However, they are not stable, and decay with rates $\gamma_4$ and
$\gamma_2+\gamma_3$ respectively. The level
scheme in the new basis is plotted in Fig. 2 b).
Here, it is evident that the system is
incoherently pumped among the driven transition $\ket{\Psi_{23}(0)}\to
\ket{\Psi_{14}(0)}$ and the two dark states. One could say that the steady
state is determined by the competition between the Hamiltonian dynamics
and the relaxation processes. Thus, some localization in one of
the dark superpositions (CPT) can occur, if this is more stable than the other,
i.e. if the rate of pumping into it is much larger than its decay rate. In
order to quantify this effect, we introduce the parameter $\alpha$ defined as:
\begin{equation}
\label{Alpha} \alpha=\frac{\gamma_4}{\gamma_2+\gamma_3}.
\end{equation}
Thus, for $\Phi=0$ and $\alpha\gg 1$ ($\alpha\ll 1$) the dark state
$|\Psi_{23}(\pi)\rangle$ ($|\Psi_{14}(\pi)\rangle$) is long lived with respect
to all other states and, at steady state, it has a high probability of
occupation. Such probability increases the more $\alpha$ differs from unity,
and approaches 1 for $\alpha\to \infty$ ($\alpha\to 0$), corresponding to the
system being trapped in $\ket{\Psi_{23}(\pi)}$ ($\ket{\Psi_{14}(\pi)}$). We
will show that due to this effect, population inversion can occur on the
transition $\ket{1}\to\ket{2},\ket{3}$ for $\alpha\gg 1$ and on
$\ket{2},\ket{3}\to\ket{4}$ for $\alpha\ll 1$.  On the other hand, such
behavior disappears as $\alpha$ approaches 1. For $\alpha=1$ and at saturation,
the system is equally scattered among all states.

\subsubsection{Comparison with the double-$\Lambda$ configuration}

In the absence of spontaneous decay, the $\diamondsuit$ configuration is
formally identical to the double-$\Lambda$ scheme, extensively studied in the
literature \cite{Kosachiov92,PhaseoniumRev,Korsunsky99}. Thus, the symmetries
induced on the coherent dynamics by the phase are exactly the same
\cite{Buckle86}. We have discussed, however, that the steady state is
critically determined by the concurrence between this symmetry and the
relaxation processes. Thus, the introduction of the spontaneous decay leads to
critical differences between the two systems. For an easier comparison we are
labelling the atomic states of the double-$\Lambda$ system as shown in the
inset of Fig.~1. In this scheme the excited states $|1\rangle$ and $|4\rangle$
decay spontaneously into the stable or metastable states $\ket{2}$ and
$\ket{3}$. In the $\diamondsuit$ scheme, the excited state $\ket{4}$ decays
into the intermediate states $\ket{2}$ and $\ket{3}$, which themselves decay
into the ground state $\ket{1}$. A first difference is that in the
$\diamondsuit$ scheme the dynamics will be phase-sensitive only when the V- or
the $\Lambda$ scheme (or both) are driven at saturation, while below saturation
it will reduce to the well-known V-configuration. In the double-$\Lambda$
system, on the other hand, phase-sensitive dynamics survives also well below
saturation \cite{PhaseoniumRev}.\\
When looking at the behavior as a function of $\Phi$, the differences are more
striking: at $\Phi=(2n+1)\pi$, for instance, the $\diamondsuit$ scheme is
pumped into the subspace $\{\ket{1},\ket{\Psi_{23}(0)}\}$, which is a closed
two-level transition, for the coherent drive as well as for the relaxation
processes. In the double-$\Lambda$ scheme, instead, the atom can be found
in any of the four states due to incoherent coupling \cite{Footnote3}.\\
At $\Phi=2n\pi$, the role of the dark states differs between the two
configurations. In the double-$\Lambda$ system CPT occurs in the state
$\ket{\Psi_{23}(\pi)}$, which is completely dark and, in the absence of other
sources of decay, stable. In this configuration, and for equal decay rates from
the excited states, the state $\ket{\Psi_{14}(\pi)}$ is never accessed. In the
$\diamondsuit$ scheme, instead, both dark states are accessed, and
(partial) CPT occurs only when their decay rates differ substantially from
one another.

\section{Steady-state solutions}

In this section we study the steady-state solution of (\ref{Master:new}) as a
function of the phase $\Phi$. We consider laser frequencies and geometries that
fulfill $\Delta k =0$ and $\Delta \omega =0,$ so that $\Phi$ does not depend on
time and space. In order to obtain simple analytic solutions we consider
resonant drives so that $\delta_j =0.$ Further, we assume that the moduli of
the Rabi frequencies are all equal, $g_{ij}=g$, and that the decay rates
fulfill the relation $\gamma_2=\gamma_3=\gamma$,
$\gamma_{42}=\gamma_{43}=\gamma_4/2$. Under these assumptions, the system
exhibits symmetry in the clockwise and anti-clockwise multiphoton excitation
paths, the difference being the phase $\Phi$.
\\ In this limit, we report and discuss the steady-state solutions of the OBE
in (\ref{Bloch1}-\ref{BlochN}) as a function of the phase $\Phi$ and of the
dimensionless parameters $\Omega=g/\gamma$ and $\alpha=\gamma_4/(2\gamma)$,
defined in (\ref{Alpha}). We remark that in the following the rotated
one-photon coherence
\begin{equation}
\tilde{\rho}_{34}= \rho_{34}e^{{\rm i}\Phi} \nonumber
\end{equation}
is reported
in the results. This frame allows to identify the real and imaginary parts of
$\rho_{12}$, $\rho_{13}$, $\rho_{24}$, $\tilde{\rho}_{34}$ with the dispersive
and absorptive response of the atomic medium to the fields which couple the
corresponding transitions \cite{QScully}.

At the end of this section we will discuss experimental situations under which
the assumptions given above are justified and discuss our results for generic
values of $\delta$, in relation to the assumption of classical motion, and
for unequal Rabi frequencies and decay rates.

\subsubsection{Case $\alpha=1$}

We first consider the case $\alpha=1$. For convenience, we separate the real
and imaginary part of the coherences, denoting them with $u_{ij}={\rm
Re}\{\rho_{ij}\}$, $v_{ij}={\rm Im}\{\rho_{ij}\}$, respectively (here,
$\tilde{u}_{34}={\rm Re}\{\tilde{\rho}_{34}\}$, $\tilde{v}_{34} ={\rm
Im}\{\tilde{\rho}_{34}\}$). The steady-state solutions of the OBE have the
form:

\begin{eqnarray}
\label{alphaeq1} \rho_{11}^{(ss)} &=&\frac{1}{D}\Bigl[1+\frac{16}{3}\Omega^2+
\frac{19}{9} \Omega^4 \left( \sin^2\frac{\Phi}{2} + 3 \right)\nonumber\\ & &
+\frac{4}{9} \Omega^6 \left( 5\sin^2\frac{\Phi}{2} + 3 \right)+
\frac{2}{9}\Omega^8\sin^2\Phi\Bigr], \\ \rho_{22}^{(ss)}=\rho_{33}^{(ss)}
&=&\frac{\Omega^2}{D}\Bigl[ 1 + \frac{1}{3} \Omega^2 \left(7 +
3\sin^2\frac{\Phi}{2}\right) \nonumber\\ & & + \frac{4}{9} \Omega^4 \left(3 +
\sin^2\frac{\Phi}{2}\right) + \frac{2}{9} \Omega^6\sin^2\Phi\Bigr], \\
\label{alphaeq1popend} \rho_{44}^{(ss)}
&=&\frac{\Omega^4}{D}\cos^2\frac{\Phi}{2} \left( 1 + \frac{4}{3} \Omega^2 +
\frac{8}{9} \Omega^4 \sin^2\frac{\Phi}{2}\right), \\ \label{alphaeq1coh}
u_{12}^{(ss)} = -u_{13}^{(ss)} &=&\frac{\Omega^3}{2D}\sin\Phi \left[ - 1 +
\frac{2}{3} \Omega^2 + \frac{8}{9} \Omega^4\sin^2\frac{\Phi}{2}\right], \\
v_{12}^{(ss)} = v_{13}^{(ss)} &=&\frac{\Omega}{D}\Bigl[ 1 + \frac{1}{3}\Omega^2
\left(7+3\sin^2\frac{\Phi}{2}\right)\nonumber\\ & & + \frac{4}{9} \Omega^4
\left(3+\sin^2\frac{\Phi}{2}\right) + \frac{2}{9} \Omega^6 \sin^2\Phi\Bigr],\\
u_{24}^{(ss)}=-\tilde{u}_{34}^{(ss)}
&=&\frac{\Omega^3}{2D}\sin\Phi\left(1+2\Omega^2+
\frac{8}{9}\Omega^4\sin^2\frac{\Phi}{2}\right),\\ \label{alphaeq1cohend}
v_{24}^{(ss)}= \tilde{v}_{34}^{(ss)} &=&
\frac{\Omega^3}{D}\cos^2\frac{\Phi}{2}\left(1+ \frac{4}{3}\Omega^2 +
\frac{8}{9} \Omega^4\sin^2\frac{\Phi}{2}\right), \\ \label{alphaeq1coh2}
u_{14}^{(ss)} &=&-\frac{\Omega^2}{D}\cos^2\frac{\Phi}{2}\left[ 1 + \frac{4}{3}
\Omega^2 + \frac{4}{9} \Omega^4 \sin^2\frac{\Phi}{2}\right],\\ v_{14}^{(ss)}
&=&\frac{\Omega^2}{2D}\sin\Phi\left[ 1 + \frac{4}{3} \Omega^2 + \frac{4}{9}
\Omega^4 \sin^2\frac{\Phi}{2}\right],\\ u_{23}^{(ss)} &=&
\frac{\Omega^2}{D}\Bigl[ 1 + \frac{2}{3}\Omega^2
\left(2+3\sin^2\frac{\Phi}{2}\right)\nonumber\\ & & + \frac{4}{9} \Omega^4
\sin^2\frac{\Phi}{2}\left(1+3\sin^2\frac{\Phi}{2}\right)\Bigr],\\
\label{alphaeq1end} v_{23}^{(ss)} &=&-\frac{\Omega^4}{D} \sin\Phi
\left[1+\frac{2}{3}\Omega^2 \sin^2\frac{\Phi}{2}\right],
\end{eqnarray}
where
\begin{eqnarray}
D &=&1 + \frac{22}{3}\Omega^2+ \frac{4}{9} \Omega^4
\left(7\sin^2\frac{\Phi}{2}+27\right) \nonumber\\ & &
+\frac{16}{9}\Omega^6\left( \sin^2\frac{\Phi}{2} + 3 \right)+ \frac{8}{9}
\Omega^8\sin^2\Phi .
\end{eqnarray}
The form of the solutions allows to identify the contributions of the various
multi-photon processes to the steady state. For instance, at second order in
$\Omega$ (i.e. at second order in $g/\gamma$) only $\rho_{14}$ depends on
the phase while the populations, one-photon coherences and $\rho_{23}$ are
independent of $\Phi$, and $\rho_{44},\rho_{24},\rho_{34},u_{12},u_{13}$, and
$v_{23}$ vanish. In fact, this limit corresponds to weak drives, and the
relevant processes consist in resonant scattering on the transitions
$\ket{1}\to\ket{2},\ket{3}$. Thus, at second order in $\Omega$ the system is
equivalent to a V configuration driven below saturation.

At higher order, the steady-state solutions are phase dependent. This is
evident, e.g., in the excited-state population, which is proportional to
$\cos^2(\Phi/2)$. In particular, at lowest order in $\Omega$,
$\rho_{44}\approx\Omega^4\cos^2(\Phi/2)$. However, as $\Omega$ is increased
this modulated dependence of the populations is lost: at leading order in
$\Omega$, and for $\Phi\neq (2n+1)\pi$, all states are equally populated. An
exceptional behavior occurs at $\Phi=(2n+1)\pi$. Here, $\rho_{44}=0$ at all
orders, while at leading order $\rho_{11}=2\rho_{22}=2\rho_{33}=1/2$. In Fig.~3
b) the populations are plotted as a function of the phase for $\Omega=2$ and
$\alpha=1$. (For comparison, Figs.~3 a) and 3 c) plot the populations for
$\alpha\ll 1$ and $\alpha\gg 1$, respectively; we will discuss these regimes in
the next subsection.) For the chosen parameters, $\rho_{11}$, $\rho_{44}$ vary
with $\Phi$, while $\rho_{22}$, $\rho_{33}$ are almost independent of the
phase.

Fig.~3 e) and 3 h) show the one-photon coherences
(\ref{alphaeq1coh}-\ref{alphaeq1cohend}) as a function of $\Phi$ for $\Omega=2$
and $\alpha=1$. Their real parts vanish for $\Phi=n\pi$, and one can easily
verify from (\ref{alphaeq1coh}-\ref{alphaeq1cohend}) that this occurs at all
orders in $\Omega$. This is a feature of resonantly-driven two-level systems,
and this result is consistent with the analysis of the previous
section. Moreover, for $\Phi=(2n+1)\pi$ one finds
$\rho_{24}=\rho_{34}=0$, which is consistent with the vanishing of
$\rho_{44}$. It is worth noting that $u_{12}$ and $u_{13}$ can show additional
zeros, as can be seen from their analytic form. These zeros depend on the
values of $\Phi$ and $\Omega$, which for $\alpha=1$ satisfy the relation
$\sin^2(\Phi/2)=3(3-2\Omega^2)/8\Omega^4$. Thus, they exist only for a certain
range of values of the Rabi frequency. Their existence can be interpreted as
interference of multi-photon scattering at all orders.\\

\indent The two-photon coherence $\rho_{14}$ is proportional to $\cos
(\Phi/2)$, a result in agreement with the calculation in (\ref{P1to4}). The
interpretation of the two-photon coherences becomes more transparent by
employing the basis of the previous section. For instance, $\rho_{14}$ can be
expressed as:
\begin{eqnarray*}
u_{14}&=&\frac{1}{2}
\left[\langle \Psi_{14}(0)|\rho|\Psi_{14}(0)\rangle-
\langle \Psi_{14}(\pi)|\rho|\Psi_{14}(\pi)\rangle\right]\\
v_{14}&=&{\rm Im}\{\langle \Psi_{14}(\pi)|\rho|\Psi_{14}(0)\rangle\} \, .
\end{eqnarray*}
Analogue equations hold for $u_{23}$ and $v_{23}$. Thus, $u_{14}=-1/2$ ($+1/2$)
corresponds to the system being in the state $\ket{\Psi_{14}(\pi)}$
($\ket{\Psi_{14}(0)}$). The imaginary part $v_{14}$ measures the coherence
between these two states. We now look at (\ref{alphaeq1coh2}-\ref{alphaeq1end})
as a function of $\Phi$, which are plotted in Fig.~3 k) and 3 n) for
$\Omega=2$. The behavior we observe is consistent with the above interpretation
in the superposition basis, and with the discussion of the populations and
one-photon coherences. At $\Phi=2n\pi$ the imaginary part $v_{14}$ ($v_{23}$)
vanishes, supporting the hypothesis of no coherence between
$\ket{\Psi_{14}(\pi)}$ and $\ket{\Psi_{14}(0)}$ ($\ket{\Psi_{23}(\pi)}$ and
$\ket{\Psi_{23}(0)}$). Moreover, $u_{14}=-u_{23}<0$, which implies, after a
straightforward calculation, that the probability to find the system in the
dark states is $1/2$. Thus, it is not proper to speak of CPT for these
parameters.\\ At leading order in $\Omega$ the coherences vanish for
$\Phi=2n\pi$, in agreement with the expectation that at saturation the system
is equally distributed between all states. At $\Phi=(2n+1)\pi$ one finds
$\rho_{14}=v_{23}=0$ while $u_{23}$ is positive and exhibits a local maximum.
This is consistent with the picture of two-level dynamics between $\ket{1}$ and
$\ket{\Psi_{23}(0)}$. \\

\indent So far we have
discussed the case $\alpha=1$, when the relaxation rates of the two-photon
coherences are the same. We have seen that the features of the phase-induced
dynamics are always recognizable in the coherences. However, at steady state
the atom is not localized in a particular atomic level or coherent
superposition of atomic levels. In general, the dependence of the populations
on the phase is washed out for increasing Rabi frequencies, except for the
vanishing of $\rho_{44}$ at $\Phi=(2n+1)\pi$. This is understood by considering
that the steady state is given by the concurrence of the coherent drive, which
has a phase-dependent symmetry, and the relaxation processes with a fixed
structure of the coupling between the atomic states. For any value of
$\Phi\neq(2n+1)\pi$, the two effects compete, and at saturation a kind of
'ergodicity' is recovered, so that the atomic states are equally
populated. On the contrary, for $\Phi=(2n+1)\pi$, an eigenspace of the coherent
scattering processes exists and is preserved by the action of the incoherent
processes. Consequently, at any value of $\Omega$ and $\alpha$ the occupation
of the state $\ket{4}$ vanishes.

\subsubsection{Case $\alpha\neq 1$}

Figures 3 a)-o) plot the steady-state solutions of the OBE for $\alpha=0.1,
1,10$. Comparing the curves, we see some general features in the behavior at
different $\alpha$. For instance, the population of the state $\ket{4}$ is
always zero for $\Phi=(2n+1)\pi$. This value of the phase is also a pole of the
coherences $\rho_{24}$, $\rho_{34}$, $\rho_{14}$, and of the real parts
$u_{12}$, $u_{23}$. Here, the population of the state $\ket{1}$ and the real
part of the two-photon coherence $u_{23}$ exhibit a local maximum. These
results are all consistent with the picture of two-level dynamics between
$\ket{1}$ and $\ket{\Psi_{23}(0)}$, as discussed in the previous section. The
steady-state values have a very transparent form for $\Phi=(2n+1)\pi$, and
read:
\begin{eqnarray}
\rho_{11}^{ss} &=&\frac{1+2 \Omega^2}{1+4\Omega^2},\\
\rho_{22}^{ss}=\rho_{33}^{ss} &=&\frac{\Omega^2}{1+4\Omega^2},\\ \rho_{44}^{ss}
&=&0,\\ v_{12}^{ss}=v_{13}^{ss} &=&\frac{\Omega}{1+4\Omega^2},\\ u_{23}^{ss}
&=&\frac{\Omega^2}{1+4\Omega^2},\\ u_{12}^{ss}
&=&u_{13}^{ss}=\rho_{24}^{ss}=\rho_{34}^{ss}=v_{23}^{ss} =\rho_{14}^{ss}=0,
\end{eqnarray}
which have been evaluated for $\Omega=g/\gamma$ and an arbitrary value of
$\alpha=\gamma_4/2\gamma$.
In these solutions, the parameter $\gamma_4$ does not
appear, showing once again that the level $\ket{4}$ does not affect the
steady-state dynamics for this value of the phase.\\
\indent A striking difference among the three regimes appears at values of the
phase close to $\Phi=2n\pi$. Here, we find population inversion on the
transitions $\ket{1}\to\ket{2},\ket{3}$ for $\alpha=10$,
($\ket{2},\ket{3}\to\ket{4}$ for $\alpha=0.1$), while the real part of the
two-photon coherence $u_{23}$ ($u_{14}$) approaches the value $-1/2$. At this
value of $\Phi$, we write the steady-state solutions as a function of $\gamma$,
$\alpha=\gamma_4/2\gamma$ and $\Omega=g/\gamma$:

\begin{eqnarray}
\rho_{11}^{ss} &=&\frac{1}{D} [\alpha^2(1+2\alpha)+\Omega^2 \alpha
(3+5\alpha+4\alpha^2) +\Omega^4(1+2\alpha)],\\ \rho_{22}^{ss}=\rho_{33}^{ss}
&=&\frac{\alpha\Omega^2}{D}[\alpha (1+2\alpha)+\Omega^2(\alpha+2)],\\
\rho_{44}^{ss} &=&\frac{\Omega^4}{D}(1+2\alpha),\\ v_{12}^{ss}=v_{13}^{ss}
&=&\frac{\alpha \Omega}{D} \left[\alpha(1+2\alpha)+\Omega^2(2+\alpha)\right]\\
v_{24}^{ss}=\tilde{v}_{34}^{ss} &=&\frac{\alpha\Omega^3}{D}(1+2\alpha)\\
u_{23}^{ss} &=&\frac{\alpha\Omega^2}{D}[\alpha(1+2\alpha)+\Omega^2(1-\alpha)]\\
u_{14}^{ss} &=&-\frac{\Omega^2}{D}[\alpha(1+2\alpha)+\Omega^2(1-\alpha)]\\
u_{12}^{ss}
&=&u_{13}^{ss}=v_{14}^{ss}=v_{23}^{ss}=u_{24}^{ss}=\tilde{u}_{34}^{ss}=0,
\end{eqnarray}
where
\begin{equation}
D=\alpha^2(1+2\alpha)+\Omega^2\alpha(3+7\alpha+8\alpha^2)
+2\Omega^4(1+4\alpha+\alpha^2).
\end{equation}
These results are plotted as a function of $\alpha$ in Fig.~4, by keeping
$\gamma$ and $\Omega$ as fixed parameters. \\ Here, we see that for $\alpha\ll
1$ the system is localized in the atomic states $\ket{1},\ket{4}$, and the
coherence $u_{14}$ has a maximum absolute value. In particular, for $\alpha\to
0$ the populations of the states $\ket{2},\ket{3}$ vanish together with the
imaginary part of all coherences. In this case the atom is in the dark state
$\ket{\Psi_{14}(\pi)}$, which is stable, and CPT occurs. Such localization
persists for small values of $\alpha$, although the populations of the
intermediate states -- and the incoherent scattering processes -- increase as
$\alpha$ approaches 1.
It is interesting that for these (small) values of $\alpha$
the system exhibits population inversion on the transitions
$\ket{2},\ket{3}\to\ket{4}$. Analogously, it can be verified that, fixed
$\gamma_4$ and $g$, for $\gamma\to 0$ the system is trapped in the dark state
$\ket{\Psi_{23}(\pi)}:$ CPT occurs in this coherence, and this implies
population inversion on the transition $\ket{1}\to \ket{2},\ket{3}$. Note that
the localization in an atomic superposition persists in the neighbourhood of
the value of the phase $\Phi=2n\pi$, as it is visible in Fig.~3 a) and c). For
instance, for $\alpha\gg 1$ population inversion occurs on the transition
$\ket{2},\ket{3}\to\ket{1}$ on an interval of values
$[2n\pi-\Phi_0,2n\pi+\Phi_0]$. The phase $\Phi_0$ satisfies the relation
$\rho_{22}(\Phi_0),\rho_{33}(\Phi_0)=\rho_{11}(\Phi_0)$, and in general
$\Phi_0$ can be said to separate two regimes, where the dynamics associated
with the symmetry at phase $2n\pi$ or with $\Phi=(2n+1)\pi$ prevails.\\
It
is interesting to note that for $\Phi=2n\pi$ the populations and in particular
the decay dependent population inversion show trends typical of  a three-level
cascade system while for $\Phi=(2n+1)\pi$ the system is effectively reduced to
a V-configuration because of $\rho_{44}=0.$

Finally, we emphasize the additional poles of the one-photon coherences which
we have identified in the analytical solutions for $\alpha=1$. We have
interpreted their origin as due to photon scattering at all orders. We remark
that they appear in $u_{12}$ and $u_{13}$ for $\alpha=0.1$, see Fig. 3 d), and
in $u_{24}$ and $\tilde{u}_{34}$ for $\alpha=10$, see Fig. 3 f).

\subsubsection{Discussion}

The analysis of this section is restricted to the choice of parameters
$\delta_j=0$, which corresponds here to values of $p_z\approx 0$. This
describes the behavior of a gas after Doppler cooling, at a thermal energy of
$\kappa_BT\sim \hbar \Gamma/2$, with $\Gamma={\rm min}(\gamma,\gamma_4)$.
Provided that the linewidth $\Gamma$ is much larger than the recoil energies,
$\Gamma\gg\hbar^2k_{ij}^2/2m$, the presented results describe sensibly the
atomic response to the drive.

When the medium is Doppler broadened, i.e. for $\kappa_BT> \hbar \Gamma/2$
(still keeping the constraint on the recoil energies), many features discussed
for the case $p_z=0$ survive and will appear in the signal measured over the
ensemble, provided that $\delta_2=\delta_3=\delta$ $\delta_4=0$, so
that two-photon transitions are Doppler-free. This situation can be realised
for degenerate intermediate-state energies ($\omega_2=\omega_3$), resonant
drives ($\omega_{12}=\omega_{13}=\omega_2-\omega_1$,
$\omega_{24}=\omega_{34}=\omega_4-\omega_2$), and laser geometries such that
the wave vectors fulfill the relation $k_{12}=k_{13}\sim -k_{24}=-k_{34}=k$.
In this way $\Phi$ does not depend on time and space
($\Delta\omega=0$, $\Delta k=0$), and $\delta$ is given by $kp_z/m$. Also in
this regime we find that for $\Phi=(2n+1)\pi$ the population of $\ket{4}$
vanishes independently of $\delta$, together with the coherences
$\rho_{24},\rho_{34}$ and $\rho_{14}$. For $\Phi=2n\pi$ and $\alpha$
sufficiently different from unity, population inversion can be observed
provided the atomic transitions are saturated \cite{Sonja}.\\
\indent Finally, we remark that only part of these considerations can be
applicable to "asymmetric" configurations, i.e. for values of the Rabi
frequencies, eigenenergies, decay rates, etc.~, which change the structure of
the Hamiltonian and relaxation processes, introducing thus either different
weights to the interfering excitation paths, and/or additional relative phases,
and/or different resonances. Here, the dependence of the steady state  on the
phase $\Phi$ cannot often be simply singled out. The dynamics is a complex
combination of all parameters, and exhibits an extremely rich variety of
phenomena, which will be subject of future investigations.

\section{Conclusion}

We have studied the dynamics of a 4-level system interacting with
lasers in a configuration which we have labelled the
$\diamondsuit$ scheme because of its geometry. This scheme has a
closed-loop excitation structure \cite{Buckle86,Kosachiov92}, i.e.
any transition amplitude between two given states is the sum of
two contributions, corresponding to two excitation paths, which
may interfere. The dynamics is determined by a large number of
parameters. Here, we have considered that both paths have the same
weight, while they differ by a relative phase $\Phi$. We have
discussed the origin of $\Phi$, and investigated the steady state
of the interacting system as a function of $\Phi$, in the regime
where the steady-state solution exists.

For the chosen parameters, the steady-state solution is phase-sensitive. This
is particularly evident in the coherences, whereas in general the phase
dependence of the population is particularly enhanced for certain ranges of
values of the relaxation rates. In particular, when the lifetimes of the
intermediate states are considerably different from the one of the upper state,
the system can exhibit population inversion for some values of the phase around
$\Phi=2n\pi$. We have interpreted and discussed this result in terms of
coherent population trapping. Nevertheless, in all regimes here considered the
population of the upper state vanishes for $\Phi=(2n+1)\pi$. We have explained
these behaviors using a convenient basis, showing that the dynamics is given by
the concurrence of the Hamiltonian evolution, which is phase-sensitive, with
the structure and non-unitarity of the relaxation processes. In particular,
for $\Phi=(2n+1)\pi$ the steady state of the system corresponds to the
steady state of a (closed) two-level transition.

The phase dependence of the Hamiltonian evolution in closed-loop schemes shares
many analogies with an atom interferometer \cite{Buckle86}. The phase
dependence survives also at steady state \cite{Kosachiov92,Korsunsky99}, and
the response of the system could be used as a device for measuring the relative
phase between laser fields. For instance, in the $\diamondsuit$ system the
phase could be measured through the population of the upper state. In fact, for
sufficiently-weak fields, the functional behavior of this population is well
approximated by $\cos^2\Phi/2$, and the fluorescence signal from the upper
state shows the features of an interference pattern which is sensitive to
$\Phi$.

This study may be useful in the spectroscopy of alkali-earth atoms, currently
investigated in experiments aiming at optical frequency standards
\cite{PTB01} or at Bose-Einstein condensation by all-optical means
\cite{Katori99}. Further, efficient laser-cooling schemes for these kind of
atoms could be developed, by exploiting the phase properties due to the atomic
motion in proper laser geometries \cite{Cooling}.

Finally, the $\diamondsuit$ scheme exemplifies a system where non-linear optics
with resonant atoms can be realized. Here, the phase is a control parameter
capable to change the response of the medium to the drive
\cite{Korsunsky99,PhaseoniumRev,Windholz96,Harris00,Windholz99,Affolderbach}.
This will be object of future investigations.

\section{Acknowledgments}

This work has been motivated by a problem posed by Robin Kaiser and
Ennio Arimondo who are kindly acknowledged. We also thank S.M. Barnett,
W. Lange, M. Scully, E. Solano, H. Walther and in particular P. Lambropoulos
for useful discussions and comments. We acknowledge support from the
TMR-networks QSTRUCT and QUANTIM, EPSRC (GR/R04096) and the Leverhulme Trust.
GLO acknowledges support from SGI.\\

\noindent $^*$ \small{Present address: University of Ulm, Abteilung
Quantenphysik, Albert-Einstein-Allee 11, D-89081 Ulm, Germany.}

\newpage

\begin{figure}[h]
\begin{center}
\epsfxsize=0.4\textwidth
\epsffile{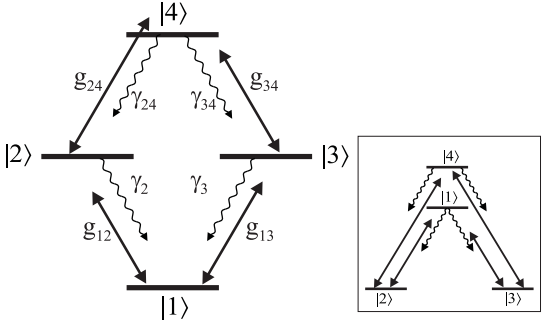}
\end{center}
\caption{Atomic level scheme and nomenclature or the relevant levels for the
$\diamondsuit$ configuration. The inset displays the Double-$\Lambda$ system
for comparison.}
\end{figure}

\begin{figure}[h]
\begin{center}
\epsfxsize=0.4\textwidth
\epsffile{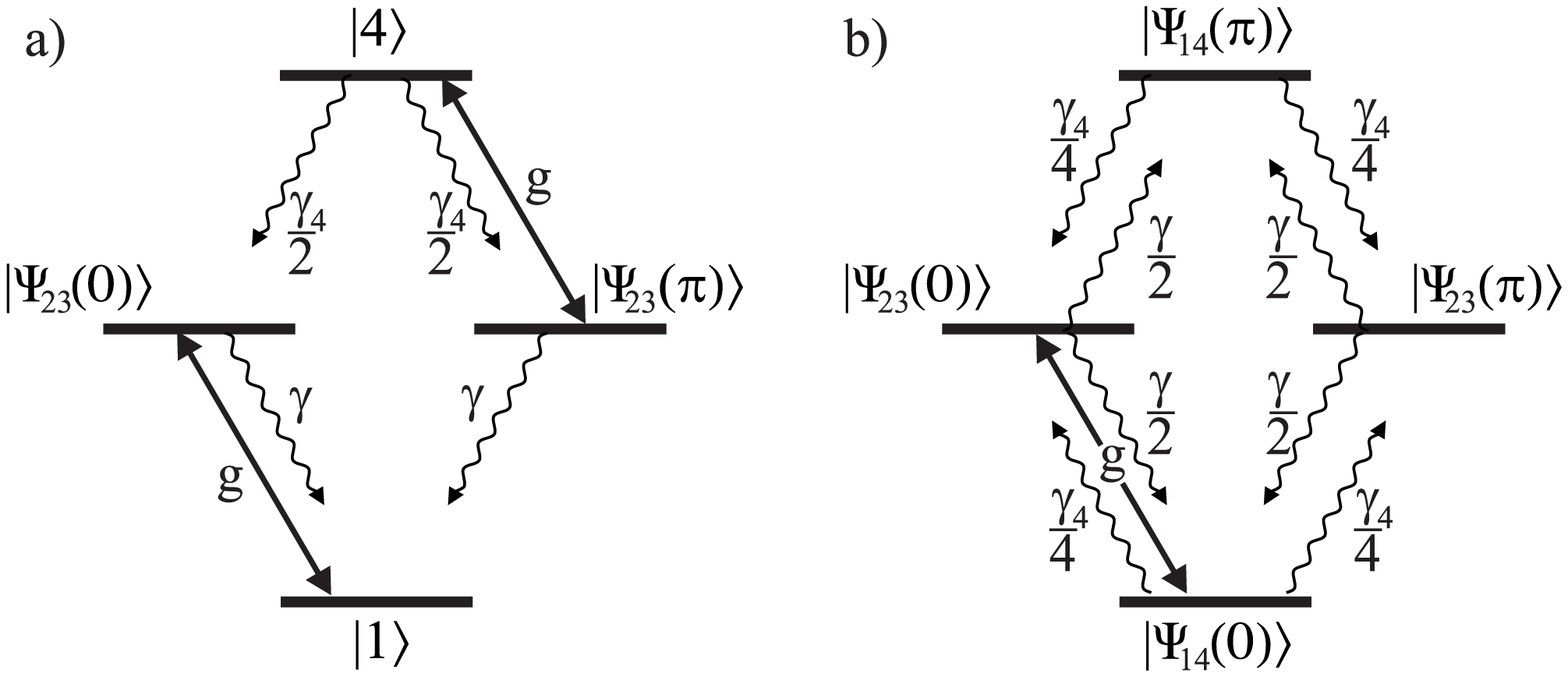}
\end{center}
\caption{Atomic level scheme in the basis of the states $\{ \ket{1},\ket{4},
\ket{\Psi_{23}(0)},\ket{\Psi_{23}(\pi)} \}$} for the cases $\Phi=(2n+1)\pi$ (a)
and $\Phi=2n\pi$ (b).
\end{figure}

\begin{figure}[h]
\begin{center}
\epsfxsize=0.8\textwidth
\epsffile{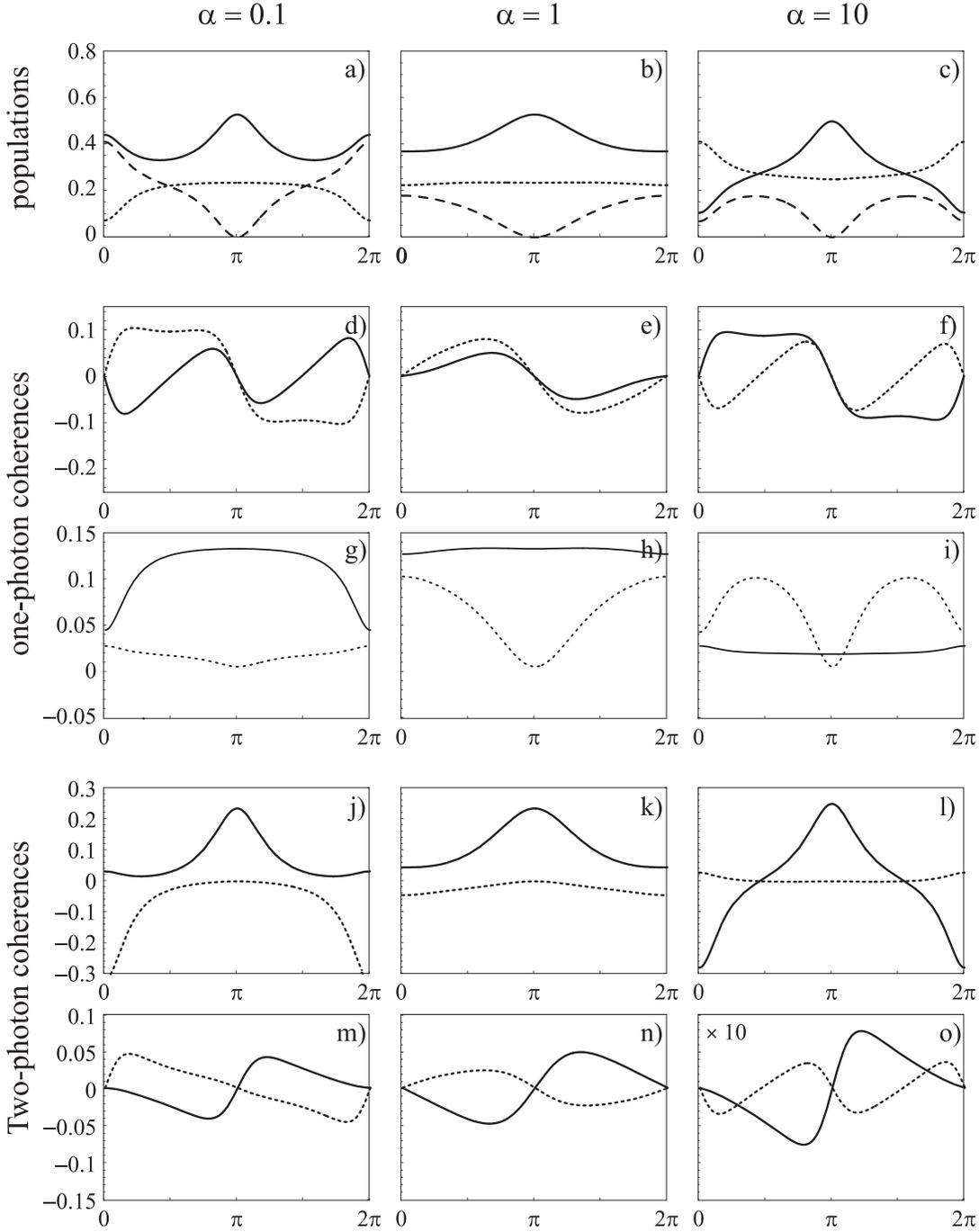}
\end{center}
\caption{Steady-state values of populations and coherences as a function of the
phase $\Phi$ for resonant drives and different decay rates: $\alpha=0.1$ with
$\gamma_4=0.2\gamma$ and $g=2 \gamma$ (left column), $\alpha=1$ with
$\gamma_4=2\gamma$ and $g=2\gamma$ (center column), $\alpha=10$ with
$\gamma_4=20\gamma$ and $g=\gamma_4$ (right column). Subplots a), b) and c):
populations of $\rho_{11}$ (solid), $\rho_{22}=\rho_{33}$ (dotted) and
$\rho_{44}$ (dashed line). Subplots d), e) and f): real part of the one-photon
coherences; $u_{12}=-u_{13}$ (solid), and $u_{24}=-\tilde{u}_{34}$ (dotted
line). Subplots g), h) and i): imaginary part of the one-photon coherences;
$v_{12}=v_{13}$ (solid), $v_{24}=\tilde{v}_{34}$ (dotted line). Subplots j), k)
and l): real part of the two-photon coherences; $u_{23}$ (solid), $u_{14}$
(dotted line). Subplot m), n) and o): imaginary part of the two-photon
coherences; $v_{23}$ (solid), $v_{14}$ (dotted line). Note the different
scaling factor of the vertical axis in subplot o).} \label{Fig3}
\end{figure}

\begin{figure}[h]
\begin{center}
\epsfxsize=0.8\textwidth \epsffile{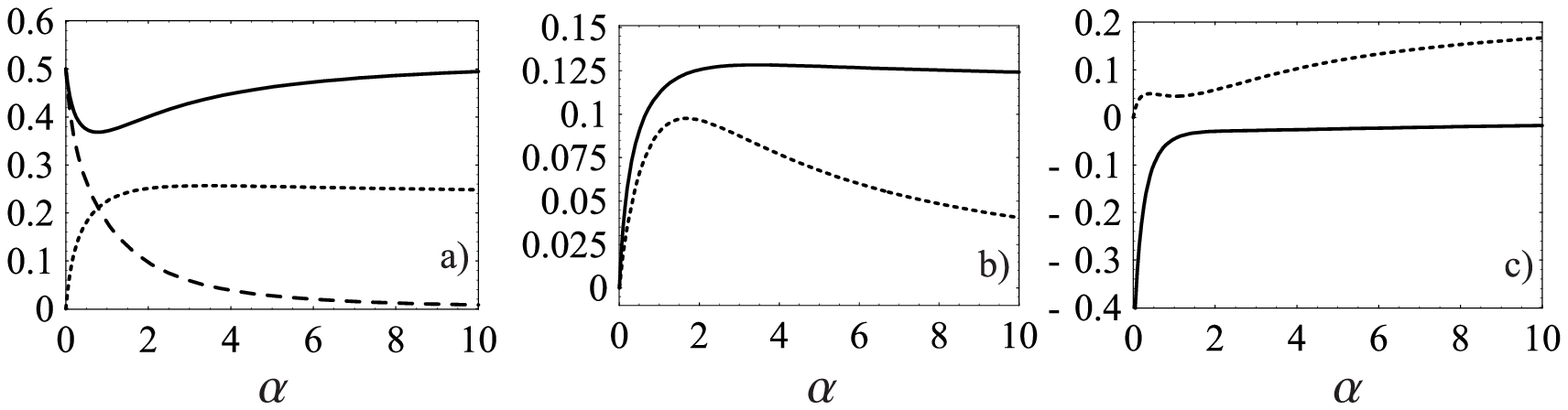}
\end{center}
\caption{Level populations and associated coherences for $\Phi=2n\pi$ as a
function of the balance of the decay rates $\alpha$ for $g=2\gamma$ (where
$\gamma$ is kept constant). (a)
populations: $\rho_{11}$ (solid line), $\rho_{22}=\rho_{33}$ (dotted line),
$\rho_{44}$ (dashed line). (b) one-photon coherences: $v_{12}=v_{13}$ (solid
line), $v_{24}=\tilde{v}_{34}$ (dotted line). (c) two-photon coherences:
$u_{23}$ (solid line), $u_{14}$ (dotted line). All other coherences vanish.}
\label{Fig4}
\end{figure}


\begin{references}
\bibitem{Footnote}
A systematic investigation should include the mechanical effects of light on
the center-of-mass motion dynamics, and the OBE should describe the coupled
dynamics of internal and external degrees of freedom. Nevertheless, in the
limit in which the internal relaxation is much faster than the rate with which
the atomic center-of-mass coordinates evolve on a macroscopic scale, one can
evaluate the internal steady state, treating these coordinates as fixed
parameters. Our analysis is restricted to this short-time behavior.
\bibitem{Footnote2}
The higher-order terms are time-dependent, and might give rise to non-trivial
effects in the long-time behavior, i.e. on a time scale longer than that on
which the steady state of the V system is approached. On such time scales the
change in time of the atomic density due, e.g., to the mechanical effects of
light should be taken into account.
\bibitem{Footnote3}
Note that for $\Phi=(2n+1)\pi$ population inversion has been predicted in the
double-$\Lambda$ configuration for different decay rates from the excited
states \cite{Kosachiov92}.
\end{references}
\end{document}